\documentclass[12pt,a4paper,final]{iopart}
\usepackage{iopams}

\expandafter\let\csname equation*\endcsname\relax
\expandafter\let\csname endequation*\endcsname\relax

\usepackage{amsmath}
\usepackage{amsfonts}
\usepackage{amsmath}
\usepackage{amssymb}
\usepackage[utf8]{inputenc}
\usepackage{graphicx}
\usepackage{bm}
\usepackage{bbm}
\usepackage{cite}

\usepackage[breaklinks=true,colorlinks=true,linkcolor=blue,urlcolor=blue,citecolor=blue]{hyperref}

\newcommand{\defeq}{\mathrel{\mathop:}=}
\newcommand{\U}{\mathcal{U}}
\renewcommand{\H}{\mathcal{H}}
\newcommand{\params}{\boldsymbol{\lambda}}

\begin{document}

\title[Reconstruction of the superstatistical temperature distribution]{Reconstruction of the superstatistical temperature distribution from single-particle kinetic energies}

\author[cor1]{Sergio Davis$^{1,2}$}
\address{$^1$Research Center in the Intersection in Plasma Physics, Matter and Complexity (P$^2$mc), Comisión Chilena de Energía Nuclear, Casilla 188-D, Santiago, Chile}
\address{$^2$Departamento de Física y Astronomía, Facultad de Ciencias Exactas, Universidad Andres Bello. Sazié 2212, piso 7, 8370136, Santiago, Chile.}
\ead{sergio.davis@cchen.cl}

\begin{abstract}
The framework of superstatistics can be used to describe non-Maxwellian single-particle velocity distributions as mixtures of Maxwellian distributions, where the weight is imposed by a probability density 
associated to the inverse temperature $\beta = 1/(k_B T)$. Among the typical model choices for $\beta$ are gamma, inverse gamma and lognormal distributions, sometimes called the universality classes of 
superstatistics. Given a set of observed velocities, however, there is no direct method to determine the underlying temperature distribution, as temperature itself is not a phase-space observable, and numerically 
inverting the Laplace transform is an unreliable process. In this work, we show that Jaynes' principle of maximum entropy can be used to successfully reconstruct the inverse temperature distribution from data, 
by using the logarithmic moments of the kinetic energy as constraints. Although the kinetic energy distributions for the three universality classes are almost indistinguishable at the same mean and variance of 
$\beta$, the method correctly discriminates between them.
\end{abstract}

\section{Introduction}

Statistical mechanical systems of classical particles in non-equilibrium steady states usually exhibit non-Maxwellian velocity distributions. A rather elegant explanation for such distributions is given 
by the theory of superstatistics~\cite{Beck2003, Beck2004}, where the non-equilibrium system is understood as being in a superposition of Maxwellian statistics.

One of the paradigmatic examples is the case of collisionless plasmas, where the mathematical structure of superstatistics appears naturally~\cite{Ourabah2015, Davis2019b}. In particular, the kappa 
distribution~\cite{Pierrard2010, Livadiotis2017, Lazar2021}, widely observed in space plasmas and originally understood~\cite{Leubner2002} in the framework of non-extensive statistical mechanics~\cite{Tsallis2009}, corresponds to superstatistics with a gamma distribution of inverse temperatures~\cite{Davis2026}. Furthermore, the other so-called universality classes, namely inverse gamma and lognormal, are also used in 
this context~\cite{Ourabah2020b, Ourabah2024}.

The inverse problem of superstatistics, that is, recovering the temperature distribution from observed particle velocities, presents itself as a major challenge. This is because, on the one hand, recovering 
the temperature distribution by numerical inversion of the Laplace transform becomes an ill-posed problem~\cite{Davies1979, Epstein2008, Marechal2024b}, and, on the other because, as was recently shown, the 
superstatistical temperature is not directly observable~\cite{Davis2018}.

In this work, we present a practical method for the reconstruction of the superstatistical inverse temperature distribution using samples of single-particle kinetic energies, method based on the application 
of Jaynes' principle of maximum entropy~\cite{Jaynes1957, Jaynes2003}. We demonstrate the accuracy of the method by reconstructing the three universality classes of superstatistics (gamma, inverse gamma and 
lognormal distributions of $\beta$) from kinetic energy data generated by each class. We also highlight the fact that these three universality classes, when imposing the same mean and variance of $\beta$, 
have almost indistinguishable kinetic energy distributions, except at the far end of the tails.

The rest of the paper is organized as follows. First, Section~\ref{sec:superstat} briefly reviews the framework of superstatistics, focusing on the description of single-particle velocity distributions.
Then, Section~\ref{sec:logmom} develops the computation of the logarithmic moments of $\beta$ for the Maxwellian distribution and also the logarithmic moments of kinetic energy, while Section~\ref{sec:maxent} 
applies the maximum entropy method constraining the logarithmic moments of kinetic energy. Section~\ref{sec:results} shows the application of this method to synthetic data from the three universality classes 
of superstatistics, and finally Section~\ref{sec:concluding} ends with some final remarks.

\section{Superstatistics and single-particle velocity distributions}
\label{sec:superstat}

Superstatistics~\cite{Beck2003, Beck2004} is a generalization of the traditional Boltzmann-Gibbs statistical mechanics, where the canonical ensemble
\begin{equation}
P(\bm \Gamma|\beta) = \frac{\exp\big(-\beta\H(\bm \Gamma)\big)}{Z(\beta)}
\end{equation}
with $\bm \Gamma$ one of the possible microstates of the system, $\beta \defeq 1/(k_B T)$ the inverse temperature, $\H(\bm \Gamma)$ the Hamiltonian and $Z(\beta)$ the partition function, is replaced by the 
joint distribution
\begin{equation}
P(\bm \Gamma, \beta|\params) = P(\beta|\params)\frac{\exp\big(-\beta \H(\bm \Gamma)\big)}{Z(\beta)}
\end{equation}
where $\params$ are the parameters of the superstatistical model, and $P(\beta|\params)$ is the superstatistical distribution of inverse temperature. This means, $\beta$ is now treated as a random variable 
coupled to $\bm \Gamma$. By integrating out the variable $\beta$ we obtain, according to the marginalization rule of probability~\cite{vonDerLinden2014},
\begin{equation}
P(\bm \Gamma|\params) = \int_0^\infty d\beta\,P(\beta|\params)\frac{\exp\big(-\beta \H(\bm \Gamma)\big)}{Z(\beta)}.
\end{equation}

\noindent
For a particle in a collisionless plasma, it has been shown~\cite{Davis2019b} that
\begin{equation}
P(\bm v|\params) = \left(\frac{m}{2\pi}\right)^{\frac{3}{2}}\int_0^\infty d\beta\,P(\beta|\params)\beta^{\frac{3}{2}}\exp\left(-\frac{\beta m\bm{v}^2}{2}\right)
\end{equation}
where
\begin{equation}
P(\bm v|\beta) = \left(\frac{m\beta}{2\pi}\right)^{\frac{3}{2}}\exp\left(-\frac{\beta m\bm{v}^2}{2}\right)
\end{equation}
is the Maxwellian distribution of velocities. The superstatistical distribution of kinetic energies is then
\begin{equation}
\begin{split}
P(k|\params) & = \left<\delta\left(\frac{m\bm{v}^2}{2}-k\right)\right>_{\params} \\
& = \left(\frac{m\beta}{2\pi}\right)^{\frac{3}{2}}\times 4\pi\int_0^\infty dv\,v^2\delta\left(\frac{mv^2}{2}-k\right)\int_0^\infty d\beta\,P(\beta|\params)\exp(-\beta k)
\end{split}
\end{equation}
which reduces to
\begin{equation}
\label{eq:probk_laplace}
P(k|\params) = \frac{2\sqrt{k}}{\sqrt{\pi}}\int_0^\infty d\beta\,\beta^{\frac{3}{2}}\,P(\beta|\params)\exp(-\beta k)
\end{equation}
that is,
\begin{equation}
\label{eq:probk_super_pre}
P(k|\params) = \int_0^\infty d\beta\,P(\beta|\params)P(k|\beta)
\end{equation}
with
\begin{equation}
\label{eq:maxwell_k}
P(k|\beta) = \frac{2}{\sqrt{\pi}}\,\beta^{\frac{3}{2}}\exp(-\beta k)\sqrt{k}
\end{equation}
the corresponding Maxwellian distribution. From \eqref{eq:probk_laplace} we can see that $P(\beta|\params)$ could, in principle, be obtained from the inverse Laplace transform of the empirical function
\begin{equation}
\hat{\rho}(k) \defeq \frac{\sqrt{\pi}\,P(k|\params)}{2\sqrt{k}},
\end{equation}
where $P(k|\params)$ is approximated by the histogram of observed samples of $k$.

\vspace{10pt}
Two invariant quantities can be defined for all superstatistical models, namely the mean inverse temperature
\begin{equation}
\beta_S \defeq \big<\beta\big>_{\params}
\end{equation}
and the inverse temperature variance 
\begin{equation}
\U \defeq \big<(\delta \beta)^2\big>_{\params}
\end{equation}

\noindent
From these two quantities, the dimensionless \emph{relative inverse temperature variance}
\begin{equation}
u \defeq \frac{\U}{(\beta_S)^2}
\end{equation}
can also be defined. Note that the limit $u \rightarrow 0$ leads to the canonical distribution (i.e. the Maxwellian in the case of velocities) for all models.

\section{Logarithmic moments of the Maxwellian distribution of energies}
\label{sec:logmom}

Let us consider the Maxwellian single-particle kinetic energy distribution in \eqref{eq:maxwell_k}. The superstatistical version of this kinetic energy distribution is given in \eqref{eq:probk_super_pre} by
\begin{equation}
\label{eq:probk_super}
P(k|\params) = \frac{2\sqrt{k}}{\sqrt{\pi}}\int_0^\infty d\beta\,P(\beta|\params)\beta^{\frac{3}{2}}\exp(-\beta k).
\end{equation}

\noindent
The $n$-th logarithmic moment of the Maxwellian in \eqref{eq:maxwell_k} is defined by
\begin{equation}
\big<(\ln k)^n\big>_\beta = \frac{2\beta^{\frac{3}{2}}}{\sqrt{\pi}}\int_0^\infty dk\,\sqrt{k}\exp(-\beta k)(\ln k)^n,
\end{equation}
and can be computed, by differentiating $n$ times the quantity
\begin{equation}
\eta(\beta; a) \defeq \int_0^\infty dk\,\exp(-\beta k)k^{\frac{a}{2}} = \Gamma\left(\frac{a}{2}+1\right)\beta^{-\frac{a}{2}-1},
\end{equation}
with respect to $a$ under the integral sign, as
\begin{equation}
\big<(\ln k)^n\big>_\beta = \left[\frac{2^n}{\eta(\beta; a)}\frac{\partial^n \eta(\beta; a)}{\partial a^n}\right]_{a = 1}.
\end{equation}

\noindent
By using the Fa\`a di Bruno formula on $\eta(\beta; a)$ written as
\begin{equation}
\eta(\beta; a) = \exp\big(\zeta(\beta; a)\big)
\end{equation}
we can, in turn, write
\begin{equation}
\frac{\partial^n}{\partial a^n}\exp\big(\zeta(\beta; a)\big) = \exp\big(\zeta(\beta; a)\big)B_n\big(\zeta', \zeta'', \ldots, \zeta^{(n)}\big)
\end{equation}
with
\begin{equation}
\zeta^{(m)}(\beta; a) = \frac{\partial^m}{\partial a^m}\ln \eta(\beta, a) = \frac{\partial^m}{\partial a^m}\left[-\frac{a+2}{2}\ln \beta + \ln \Gamma\Big(\frac{a}{2}+1\Big)\right]
\end{equation}
therefore
\begin{equation}
\zeta^{(m)}(\beta; a) = \begin{cases}
-\frac{1}{2}\ln \beta + \frac{1}{2}\psi\Big(\frac{a}{2}+1\Big), \;\;\text{for}\;\;m = 1,\\[10pt]
2^{-m}\psi^{(m-1)}\Big(\frac{a}{2}+1\Big), \;\;\text{for}\;\;m > 1.
\end{cases}
\end{equation}
with $\psi^{(m)}$ the $m$-th derivative of the digamma function, and we have
\begin{equation}
\label{eq:logmom_bell}
\big<(\ln k)^n\big>_\beta = \sum_{j=1}^n B_{n, j}\big(f_0 - \ln \beta, f_1, f_2, \ldots, f_{n-j}\big)
\end{equation}
where $f_0, f_1, \ldots, f_{n-1}$ are universal constants defined by
\begin{equation}
f_m \defeq \psi^{(m)}\Big(\frac{3}{2}\Big).
\end{equation}
The right-hand side of \eqref{eq:logmom_bell} is, in fact, a polynomial in $\ln \beta$, which can be explicitly written as
\begin{equation}
\label{eq:logmom_beta}
\big<(\ln k)^n\big>_\beta = C_n + \sum_{j=1}^n (-1)^j\binom{n}{j}C_{n-j}\,(\ln \beta)^j
\end{equation}
with $C_0 = 1$ and
\begin{equation}
C_m \defeq \sum_{j=1}^m B_{m, j}\big(f_0, f_1, \ldots, f_{m-j}\big)
\end{equation}
for $m > 1$. 

\begin{table}
\begin{center}
\begin{tabular}{|c|c|}
\hline
$m$ & $C_m$ \\
\hline
0 & 1 \\
1 & 0.03648997398 \\
2 & 0.9361337187 \\
3 & -0.7264153333 \\
4 & 3.917155434 \\
5 & 10.49704766 \\
6 & 47.61694498 \\
\hline
\end{tabular}
\end{center}
\caption{Numerical values of the coefficients $C_m$ for $m=0,1,2,3,4,5,6$.}
\end{table}

\section{The maximum entropy method with logarithmic-moment constraints}
\label{sec:maxent}

Now, let us consider a maximum entropy problem with $N$ constraints, of the form
\begin{equation}
\label{eq:constraint}
\big<(\ln k)^n\big>_{\params} = L_n,
\end{equation}
for $n=1,2,\ldots, N$, where the values $L_n$ are known. Writing the constraint in \eqref{eq:constraint} as
\begin{equation}
\Big<\big<(\ln k)^n\big>_\beta\Big>_{\params} = L_n,
\end{equation}
and using \eqref{eq:logmom_beta}, we have that, maximizing the relative entropy
\begin{equation}
\mathcal{S}[p] \defeq -\int_0^\infty d\beta\,p(\beta)\ln \frac{p(\beta)}{P(\beta|\varnothing)},
\end{equation}
subject to the constraints in \eqref{eq:constraint}, where $P(\beta|\varnothing)$ is a prior distribution for $\beta$, the maximum entropy superstatistical model is
\begin{equation}
P(\beta|\params) = \frac{1}{Z(\params)}P(\beta|\varnothing)\exp\left(-\sum_{n=1}^N\lambda_n (\ln \beta)^n\right),
\end{equation}
where the value of the Lagrange multipliers $\params$ is given by the solution of the non-linear system
\begin{equation}
L_n = C_n + \sum_{j=1}^n (-1)^j\binom{n}{j}C_{n-j}\,\big<(\ln \beta)^j\big>_{\params},
\end{equation}
which can be written in terms of derivatives of the partition function as
\begin{equation}
L_n = C_n - \sum_{j=1}^n (-1)^j\binom{n}{j}C_{n-j}\,\frac{\partial}{\partial \lambda_j}\ln Z(\params).
\end{equation}

Once the optimal parameters $\params$ are found, the reconstructed superstatistical model for $k$ is
\begin{equation}
P(k|\params) = \frac{2\sqrt{k}}{\sqrt{\pi}\,Z(\params)}\int_0^\infty d\beta\,P(\beta|\varnothing)\beta^{\frac{3}{2}}\exp\left(-\beta k -\sum_{n=1}^N \lambda_n (\ln \beta)^n\right).
\end{equation}

\section{Reconstruction of the universality classes from kinetic energy samples}
\label{sec:results}

In the following, we will show the reconstruction of each of the three universality classes of superstatistics, namely gamma ($G$), inverse gamma ($IG$) and lognormal ($L$), with datasets 
generated using so-called \emph{ancestral sampling}~\cite{Barber2012}, with 4$\times$10$^6$ samples each, at $u$ = 0.25 and $\beta_S$ = 1.25. In this method of sampling, instead of sampling from the 
marginal distribution of kinetic energy $P(k|\params)$, we obtain a sample $\beta_i$ from the superstatistical inverse temperature distribution $P(\beta|\params)$, and then a sample $k_i$ from the 
canonical (Maxwellian) conditional distribution $P(k|\beta)$ in \eqref{eq:maxwell_k}. The successive pairs $(\beta_1, k_1), (\beta_2, k_2), \ldots$ are genuine samples from the joint distribution 
\begin{equation}
P(\beta, k|\params) = P(k|\beta)P(\beta|\params),
\end{equation}
and therefore the samples $(k_1, k_2, \ldots)$ are appropriate samples from the marginal $P(k|\params)$. 

\vspace{10pt}
\noindent
For the gamma superstatistics, the probability density for $\beta$ can be written, in terms of $u$ and $\beta_S$, as~\cite{Davis2023e}
\begin{equation}
P(\beta|G, u, \beta_S) = \frac{1}{u\beta_S\,\Gamma\Big(\frac{1}{u}\Big)}\exp\left(-\frac{\beta}{u\beta_S}\right)\left(\frac{\beta}{u\beta_S}\right)^{\frac{1}{u}-1}
\end{equation}
and, accordingly, by the use of \eqref{eq:probk_super} we obtain the marginal distribution for the single-particle kinetic energy as
\begin{equation}
P(k|G, u, \beta_S) = \frac{2(u\beta_S)^{\frac{3}{2}}\,\Gamma\Big(\frac{3}{2}+\frac{1}{u}\Big)}{\sqrt{\pi}\,\Gamma\Big(\frac{1}{u}\Big)}\Big[1 + u\beta_S k\Big]^{-\left(\frac{1}{u}+\frac{3}{2}\right)}\sqrt{k},
\end{equation}
corresponding to the kappa distribution of velocities. Similarly, for the inverse gamma superstatistics we have that the probability density of $\beta$ is, in terms of $u$ and $\beta_S$, given by
\begin{equation}
P(\beta|IG, u, \beta_S) =\frac{u}{(1 + u)\beta_S\,\Gamma\Big(\frac{1}{u}+2\Big)}\left(\frac{\beta_S(1 + u)}{u\beta}\right)^{\frac{1}{u}+3}\exp\left(-\frac{\beta_S(1 + u)}{u\beta}\right).
\end{equation}

Again, by using \eqref{eq:probk_super}, the corresponding distribution of single-particle kinetic energies is given in closed form as
\begin{equation}
P(k|IG, u, \beta_S) = \frac{4\,k^{\frac{1}{u}+\frac{5}{2}}}{\sqrt{\pi}\,\Gamma\Big(\frac{1}{u}+2\Big)}K\left(\frac{1}{2}+\frac{1}{u}, 2\sqrt{\frac{(1+u)\beta_S k}{u}}\right)
\left[\frac{\beta_S(1+u)}{uk}\right]^{\frac{7}{4}+\frac{1}{2u}},
\end{equation}
where $K_\nu(z)$ is the modified Bessel function of the second kind. Lastly, for the lognormal superstatistics, the probability density of $\beta$ in terms of $u$ and $\beta_S$ is
\begin{equation}
P(\beta|L, u, \beta_S) = \frac{1}{\sqrt{2\pi\,\ln\,(1+u)}\beta}\exp\left(-\frac{1}{2\ln\,(1+u)}\left(\ln \left[\frac{\beta\sqrt{1+u}}{\beta_S}\right]\right)^2\right),
\end{equation}
but, unlike the previous cases, there is no closed form for the superstatistical distribution $P(k|L, u, \beta_S)$. Nevertheless, it can be computed numerically by using \eqref{eq:probk_super}.

\begin{figure}[h!]
\begin{center}
\includegraphics[width=0.49\textwidth]{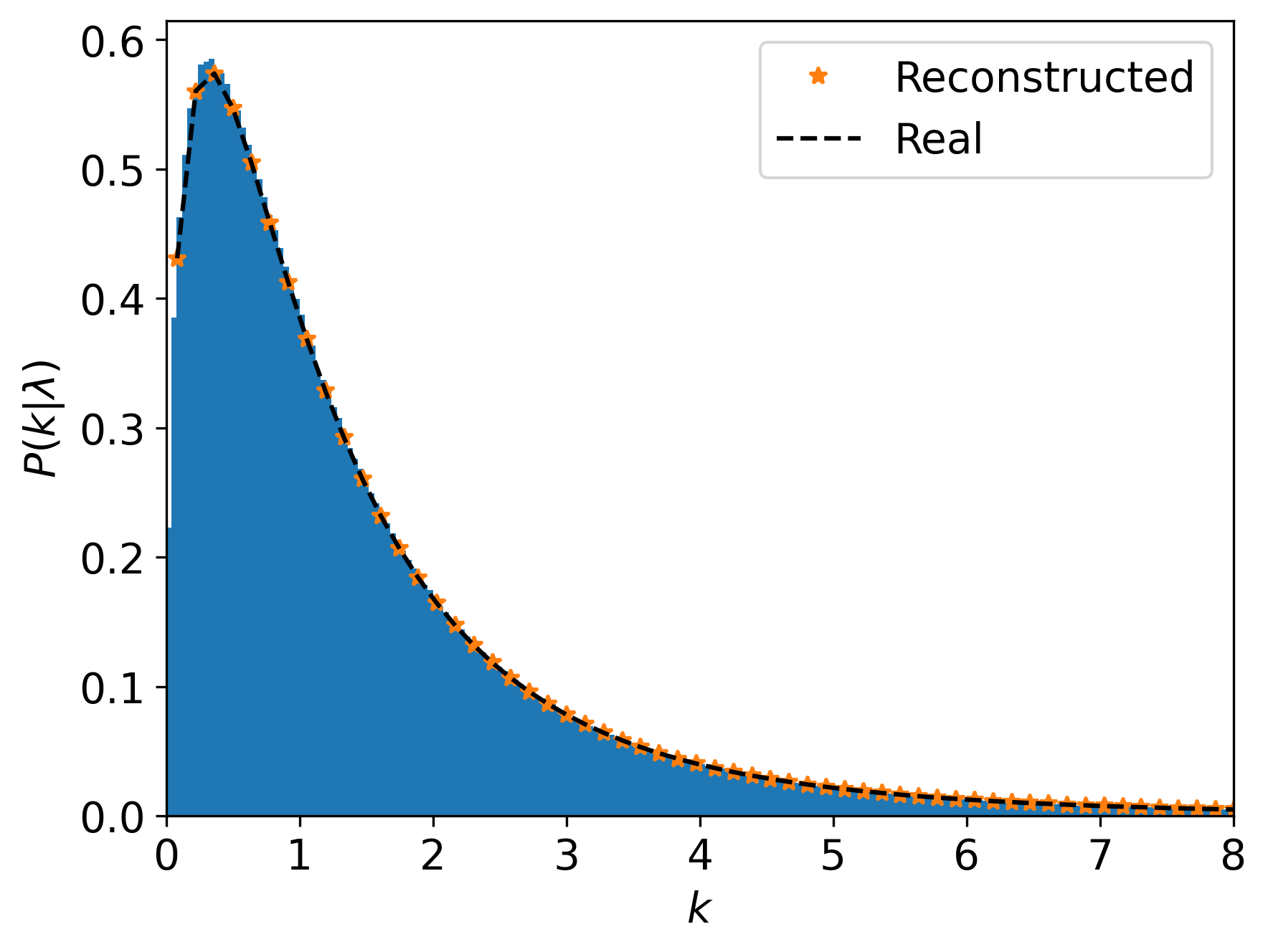}
\includegraphics[width=0.49\textwidth]{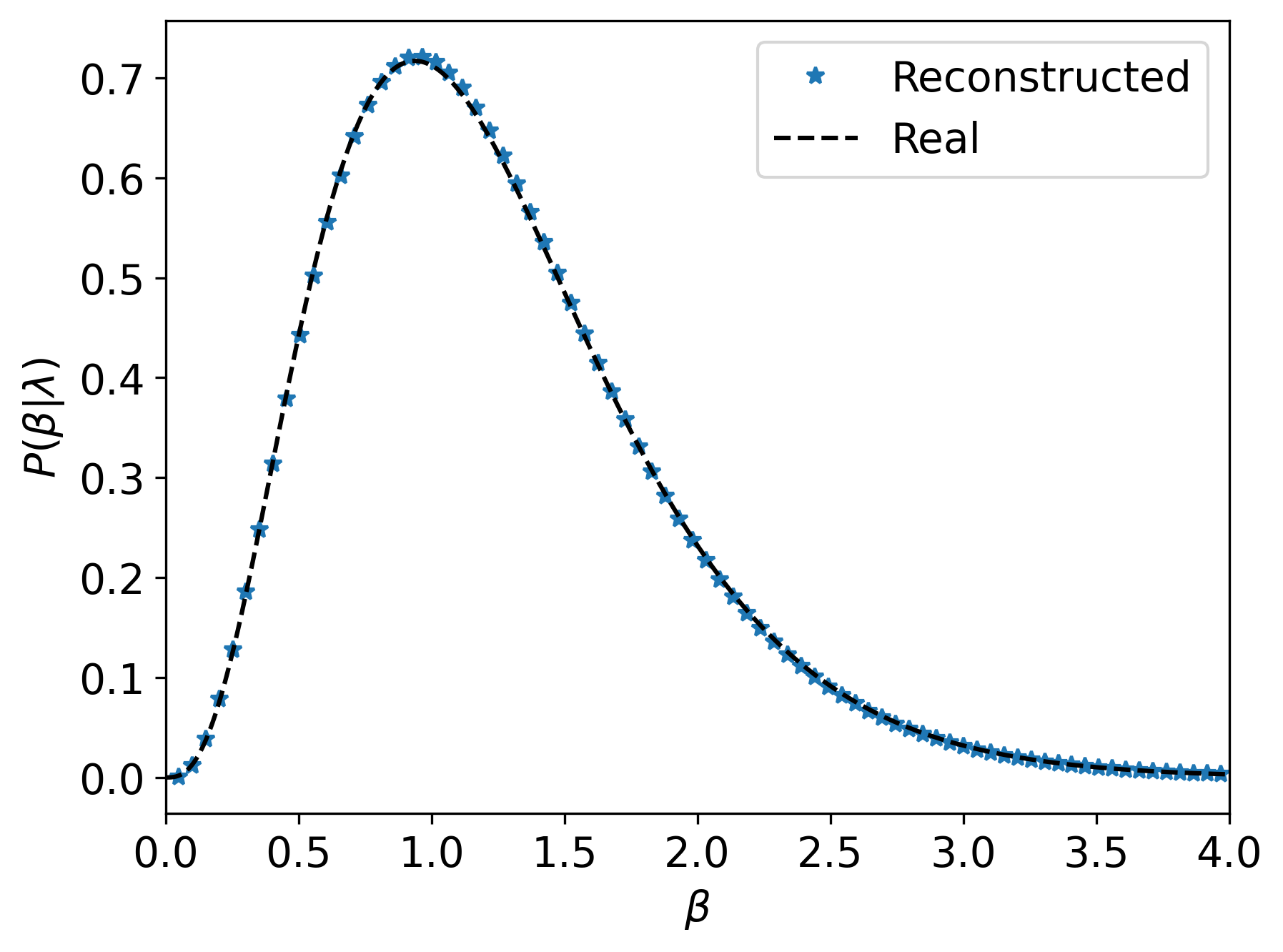}
\includegraphics[width=0.49\textwidth]{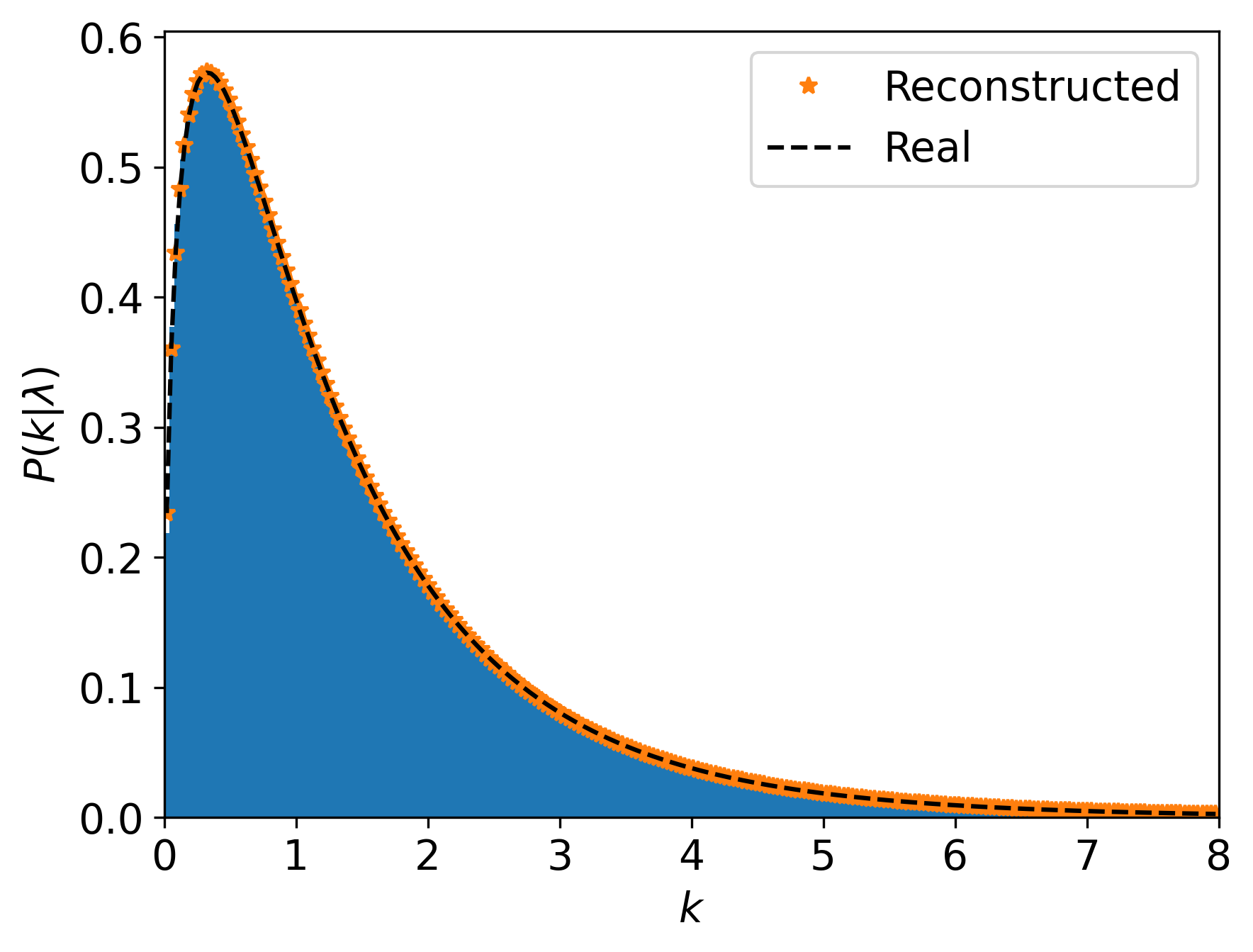}
\includegraphics[width=0.49\textwidth]{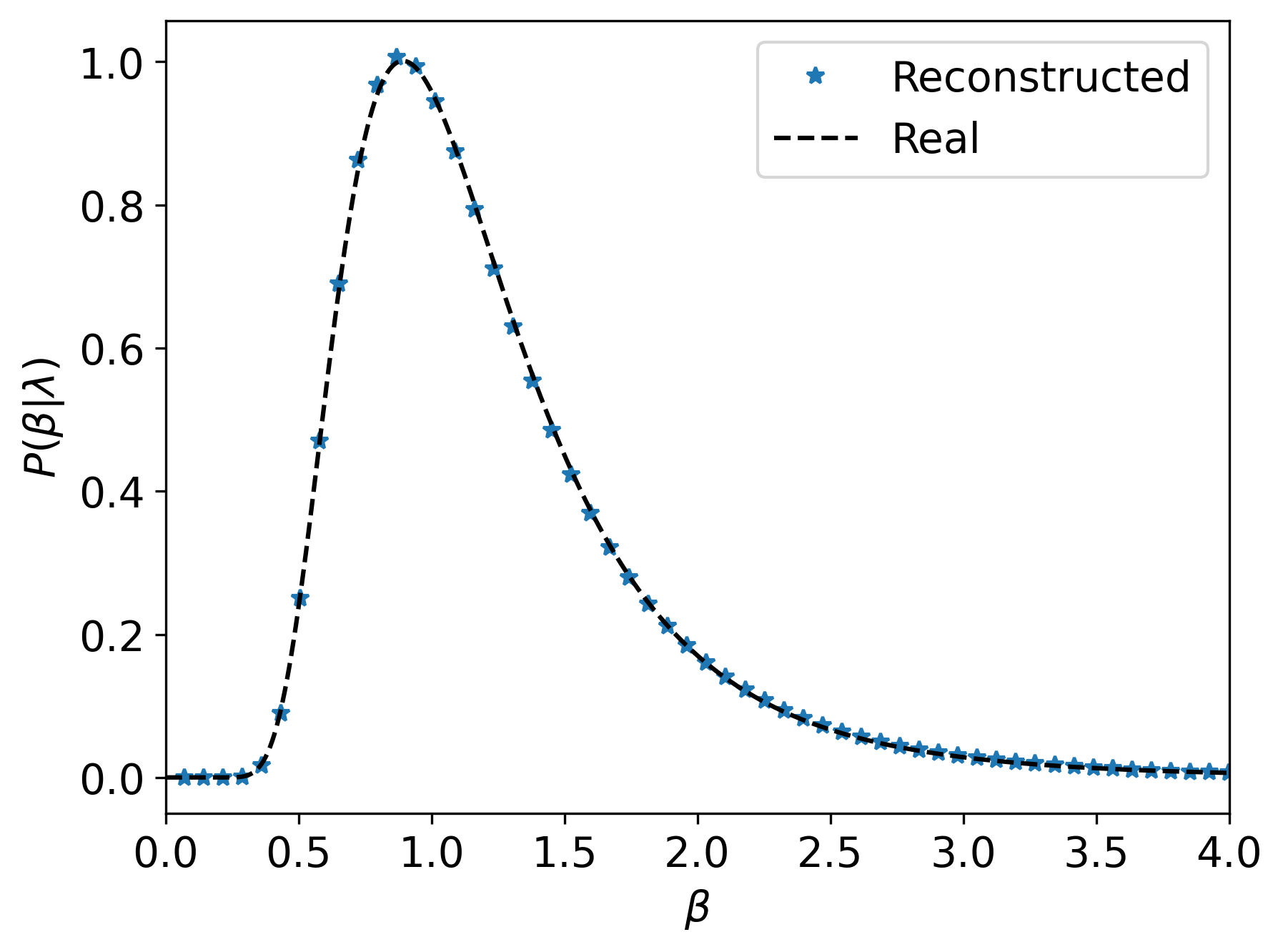}
\includegraphics[width=0.49\textwidth]{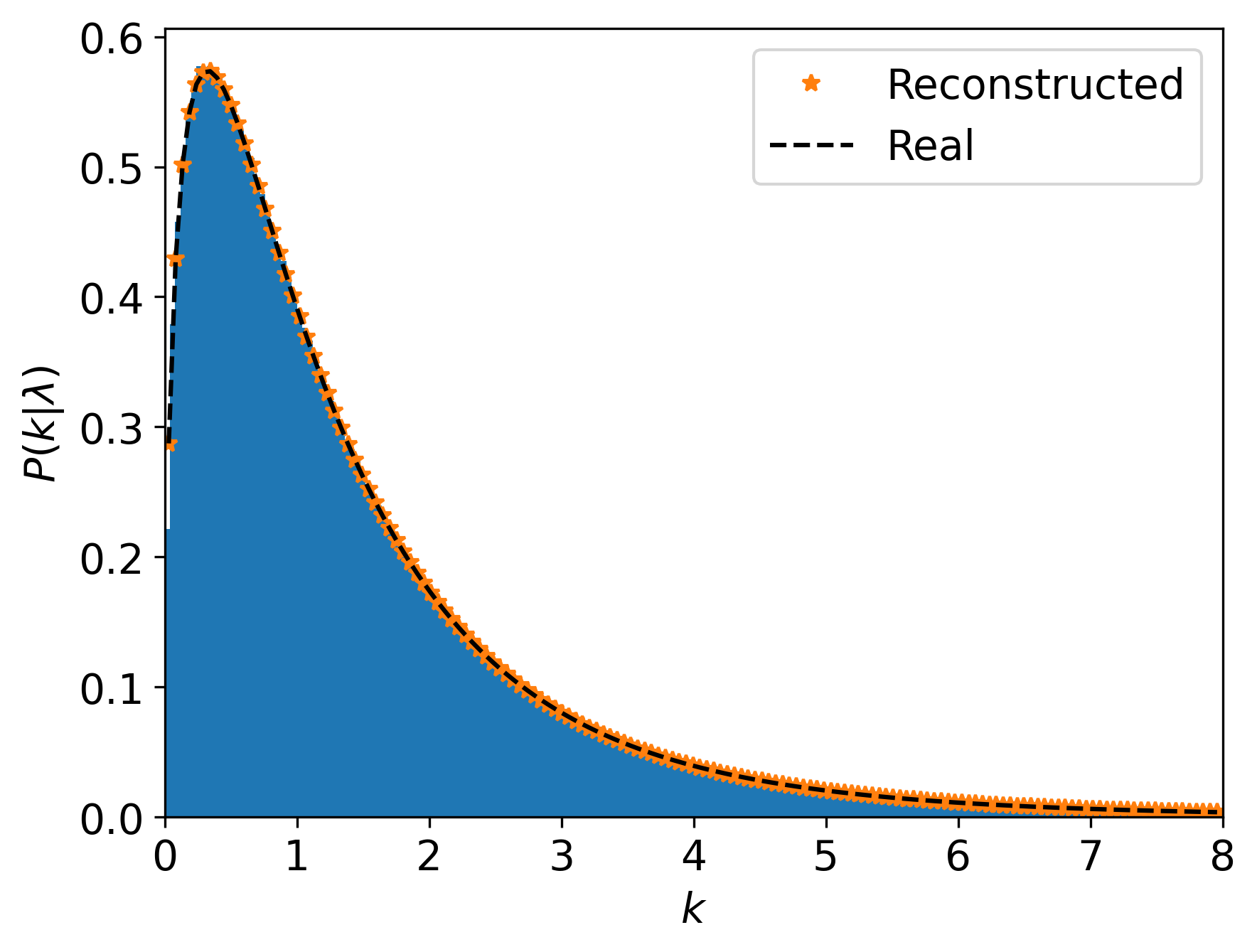}
\includegraphics[width=0.49\textwidth]{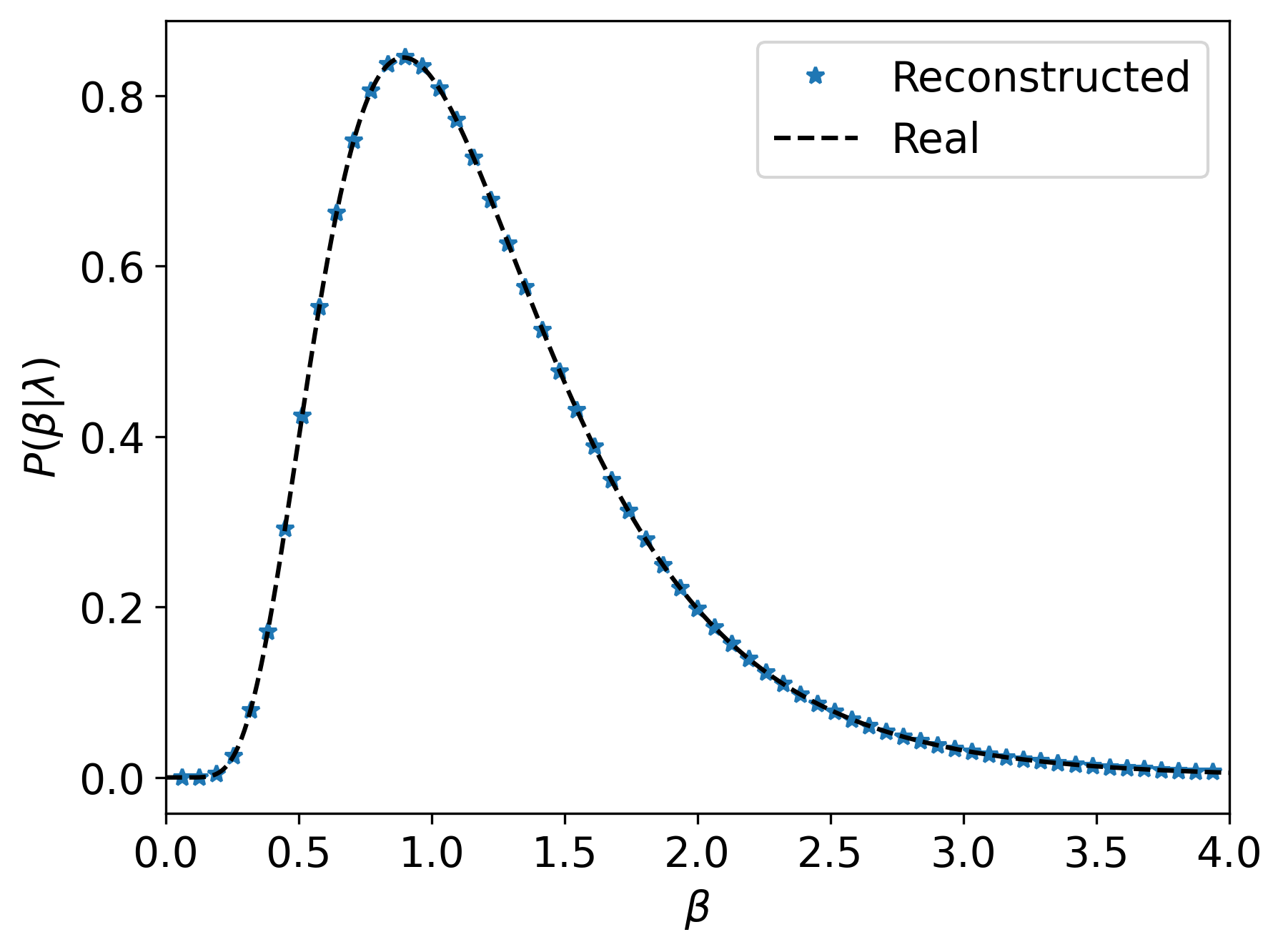}
\end{center}
\caption{Reconstruction of the inverse temperature distribution for the three universality classes of superstatistics at $u$ = 0.25 and $\beta_S$ = 1.25. Upper row, gamma distribution ($N$ = 4); middle row, 
inverse gamma distribution ($N$ = 4); lower row, lognormal distribution ($N$ = 2).}
\label{fig:reconstruct}
\end{figure}

\begin{figure}[h!]
\begin{center}
\includegraphics[width=0.49\textwidth]{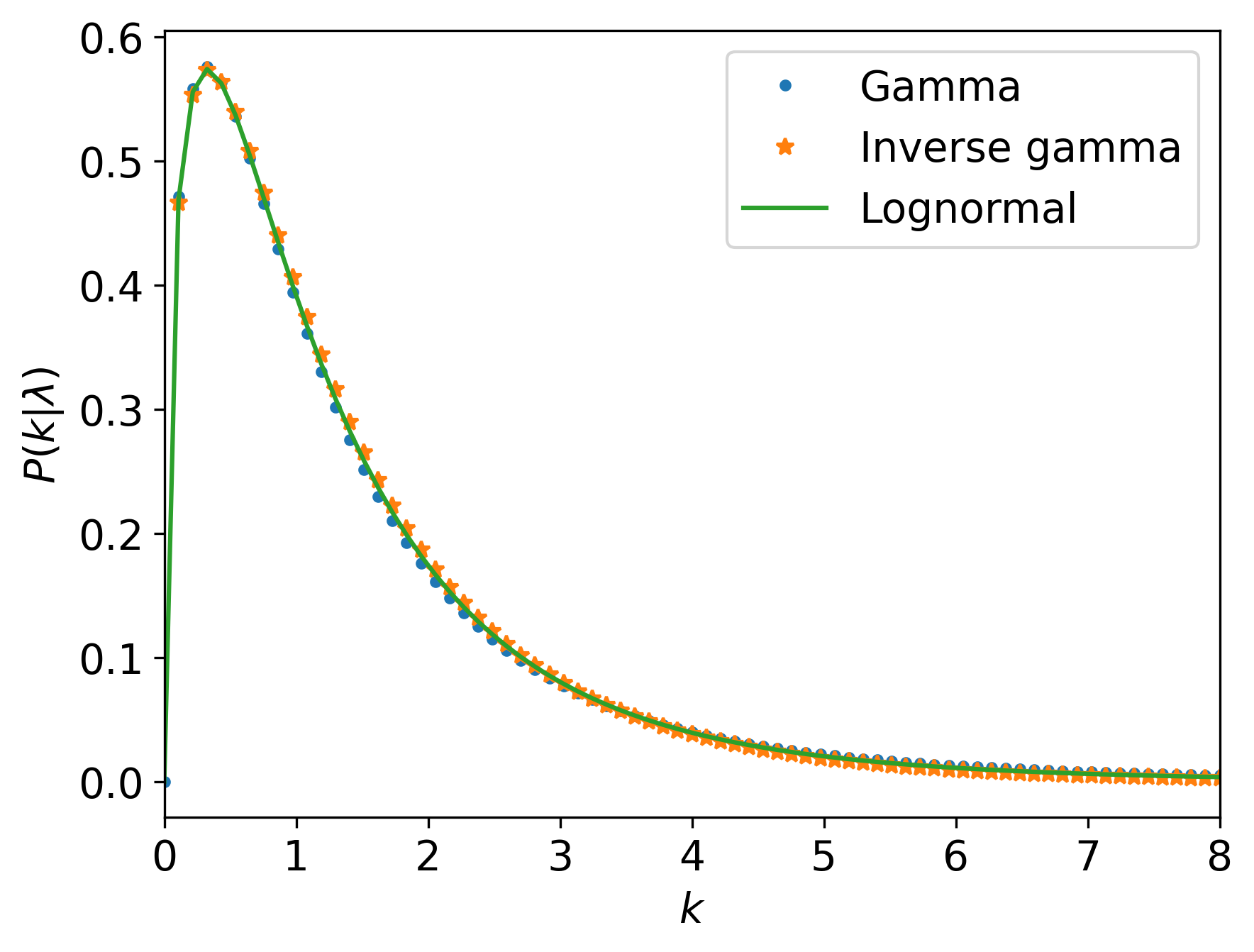}
\includegraphics[width=0.49\textwidth]{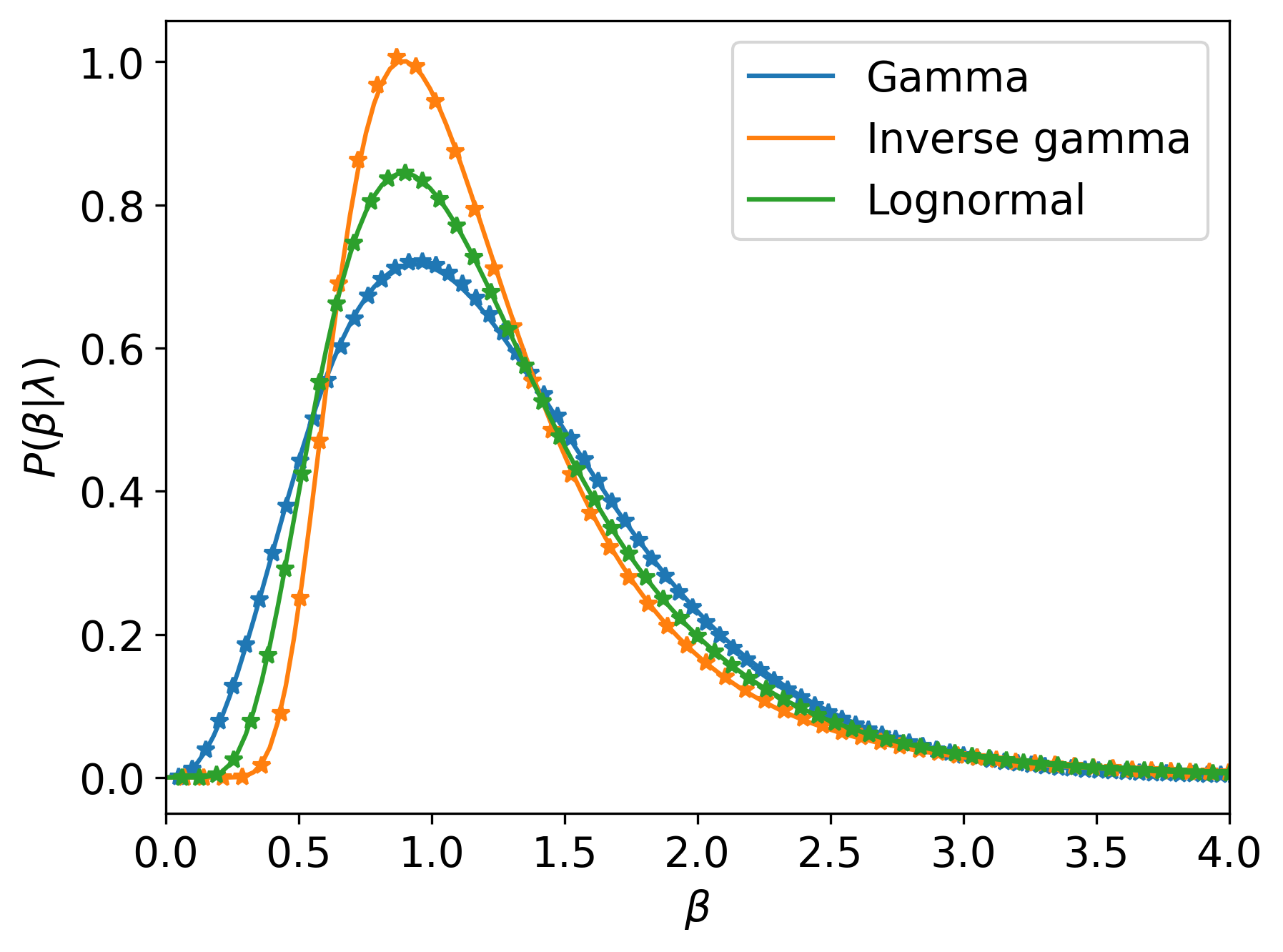}
\includegraphics[width=0.49\textwidth]{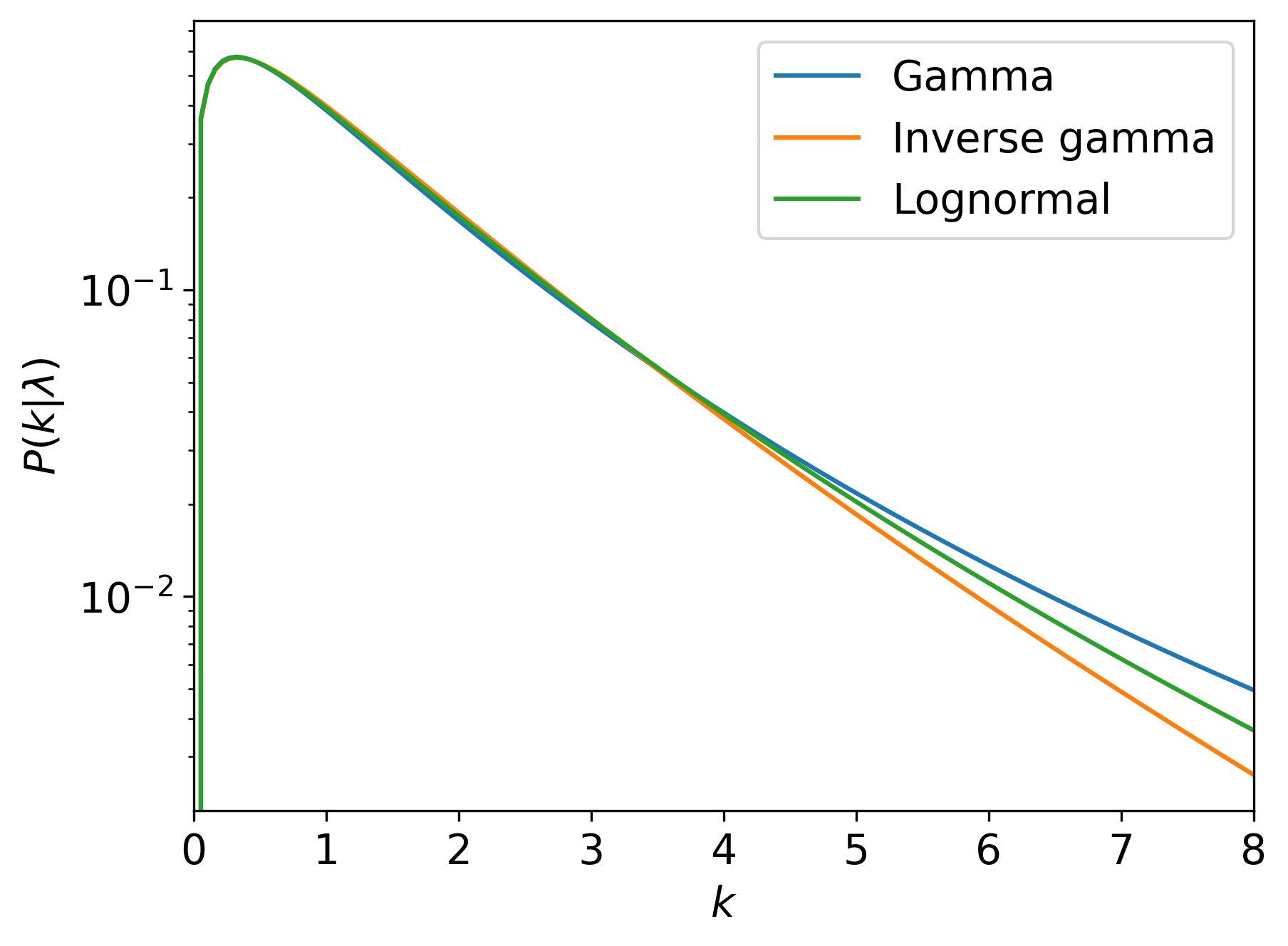}
\includegraphics[width=0.49\textwidth]{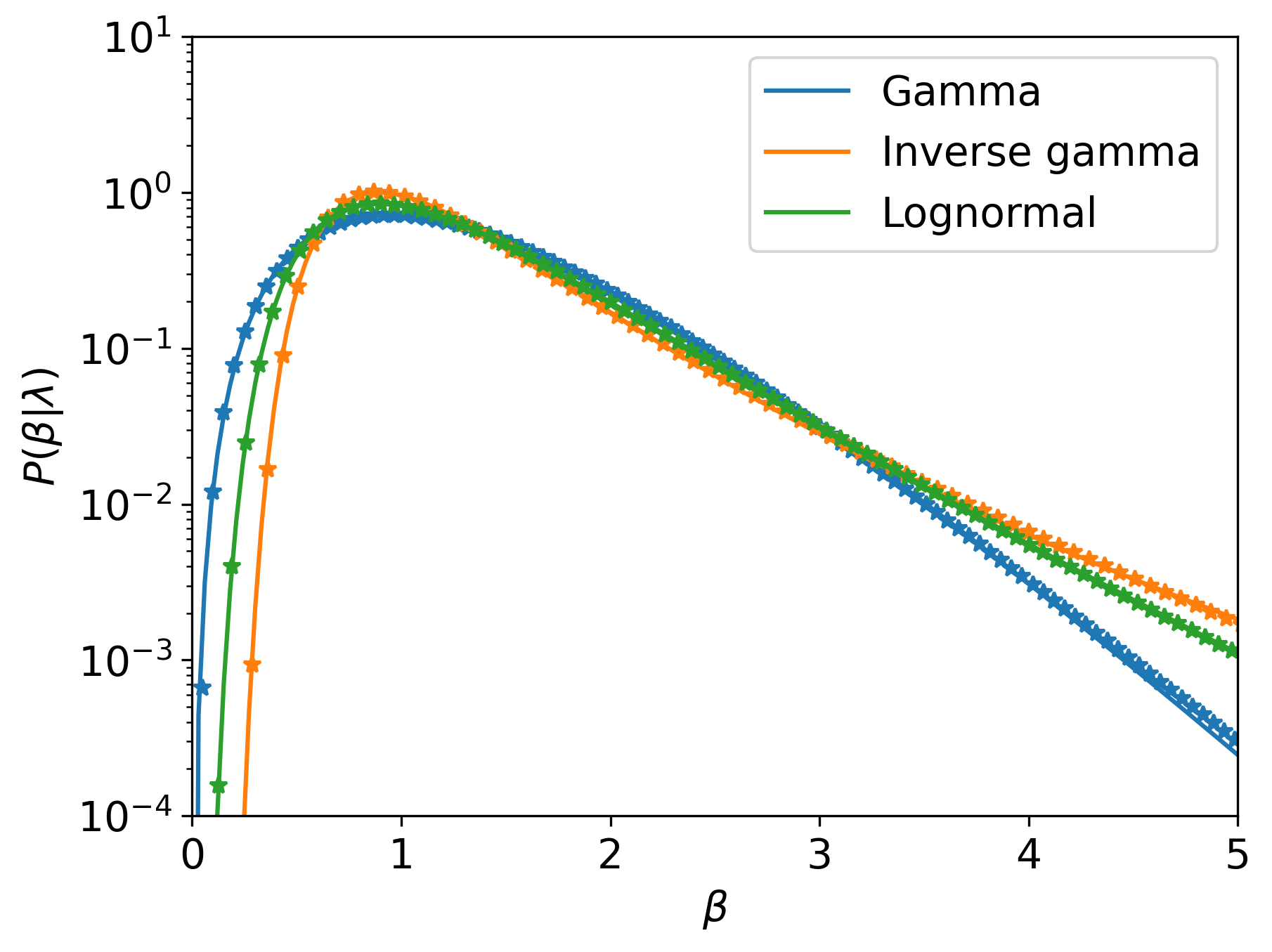}
\end{center}
\caption{Comparison of the three universality classes for $u$ = 0.25 and $\beta_S$ = 1.25. Upper panels, normal scale; lower panels, logarithmic scale.}
\label{fig:comp}
\end{figure}

Figure~\ref{fig:reconstruct} shows the reconstructed inverse temperature distributions (right column) and the reconstructed single-particle kinetic energy distributions (left column), together with 
their corresponding kinetic energy histograms, with fixed $u$ = 0.25 and $\beta_S$ = 1.25. In all cases the reconstruction is highly accurate, and we can also see that the kinetic energy distributions 
are almost identical for the three universality classes while fixing $u$ and $\beta_S$. A direct comparison between the three models is shown in Figure~\ref{fig:comp}, where we can see that only in 
logarithmic scale the difference between the three kinetic energy distributions is apparent, at the far end of the tails. Table~\ref{tbl:betamoments} shows the inferred values of $u$ and $\beta_S$ 
for the three models, together with their percent error.

\begin{figure}[h!]
\begin{center}
\includegraphics[width=0.6\textwidth]{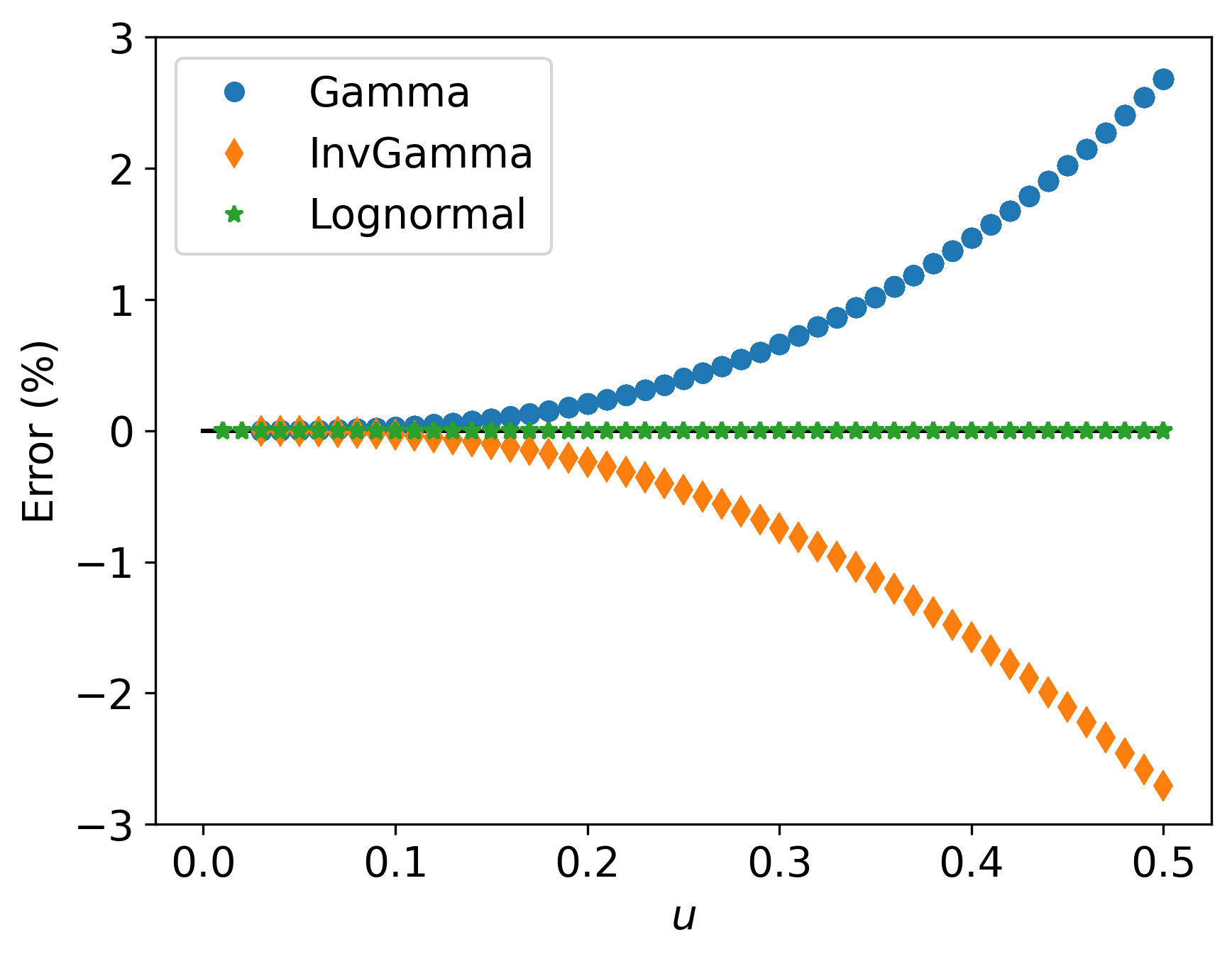}
\end{center}
\caption{Worst-case signed relative errors when estimating $u$ for the three universality classes. For each value of $u$, values of $\beta_S$ between 0.1 and 3 were considered, with $N$ = 4 logarithmic constraints.}
\label{fig:errors}
\end{figure}

Finally, Figure~\ref{fig:errors} shows the worst-case, signed relative errors for the three universality classes as functions of $u$, for values of $\beta_S$ between 0.1 and 3 and reconstructed using 
$N$ = 4 logarithmic constraints. The error increases with $u$, but it always remains below 3\%. Interestingly, for the gamma distribution the value of $u$ is overestimated, while for the inverse gamma its value is underestimated.

\begin{table}
\begin{center}
\begin{tabular}{|c|c|c|}
\hline
Model & $u$ & $\beta_S$ \\
\hline
Gamma         & 0.24788 (0.85\%) & 1.24858 (0.11\%) \\
Inverse gamma & 0.24019 (3.92\%) & 1.24950 (0.04\%) \\
Log-normal    & 0.24717 (1.13\%) & 1.24926 (0.06\%) \\
\hline
\end{tabular}
\end{center}
\caption{Inferred values of $u$ and $\beta_S$ for the three universality classes, with original values $u$ = 0.25 and $\beta_S$ = 1.25. Percent errors are shown in parentheses.}
\label{tbl:betamoments}
\end{table}

\newpage
\section{Concluding remarks}
\label{sec:concluding}

We have shown that a practical application of Jaynes' maximum entropy principle allows for the accurate reconstruction of the inverse temperature distribution underlying single-particle kinetic energies of a 
superstatistical system. 

All three major superstatistical families, i.e. the universality classes (gamma, inverse gamma and lognormal) are correctly reconstructed for values of $u$ between 0 and 1/2, with percent errors in $u$ below 
3\%. A novel finding of this work is that single-particle kinetic energy distributions obtained from these three families at the same value of $u$ and $\beta_S$ are strikingly similar, which suggests that 
kappa-like distributions may be found in contexts radically different from space plasmas.

\section*{Acknowledgments}

Numerical computations were supported by the NLHPC (ECM-02), Chile.

\newpage
\section*{References}

\bibliography{logmom}
\bibliographystyle{unsrt}

\end{document}